\documentclass{article}
\usepackage{amsmath}

\newtheorem{theorem}{Theorem}
\newtheorem{acknowledgement}[theorem]{Acknowledgement}

\input{tcilatex}

\begin{document}

\title{Simple system to measure the Earth's magnetic field}
\author{R. Akoglu$^{\ast }$, M. Halilsoy$^{\dagger }$ and S. Habib
Mazharimousavi$^{\ddag }$ \\
Department of Physics, Eastern Mediterranean University,\\
G. Magosa, north Cyprus, Mersin-10, Turkey\\
}
\maketitle

Our aim in this proposal is by using the Faraday's law of induction as a
simple lecture demonstration to measure the Earth's magnetic field ($%
\overrightarrow{B}$). This will also enable the students to learn about how
electric power is generated from the rotational motion. Obviously the idea
is not original, yet it may be attractive in the sense that no sophisticated
devices are used. All equipment that we need are available in an elementary
physics laboratory which are displayed in Fig. 1. The square, wooden coil
and handmade belt system to rotate the coil may require some craftsmanship,
which once made, can be used for years. First of all by using a compass we
orient the table parallel to the direction of the Earth's horizontal
component of $\overrightarrow{B}$ field. This is necessary to maximize the
Earth's field which can suppress the noise effects as much as possible. It
is preferable to minimize also any environmental effects by conducting the
experiment away from power lines, if possible of course.

The induced $emf$ according to the law of induction is given by\cite{1,2} 
\begin{equation}
\varepsilon =-N\frac{d\phi }{dt}
\end{equation}%
where $N=$ number of turns in the coil and $\phi $ is the magnetic flux
which is changing in time due to rotation.

\textbf{Apparatus and Experiment}

1) A square shaped wooden block of area $0.25m^{2}$ to serve as coil having $%
N=100$ turns of copper wire (with $1.1mm$ in diameter).

2) A hand-driven mechanical system connected through a belt to the wooden
coil. The handle is mounted on an empty wooden box fixed appropriately on
the table. (Remark: A car's wiper motor also can serve even better, to
rotate the coil. Since we conducted the experiment in both ways we reached
the conclusion that a wiper motor gives more efficient results).

3) A $200mV$ voltmeter (or digital avometer) and a rheostat with their
cables.

4) A compass to align the experiment table parallel to the Earth's
horizontal component of $\overrightarrow{B}$ field. As the coil rotates in
this particular case its axis remains perpendicular to $\overrightarrow{B}$.

The experiment table is shown in Fig 1. This shows also the the brush system
which collects the generated alternating current (voltage) from the rotating
coil. As remarked before, we set the coil in rotation, either manual or by
getting power from a wiper motor. For each 10 rotations \ \ we read the $emf$
from the voltmeter and prepare the Tab. 1. Then, the amplitude of $emf$
versus the inverse of the period $T$ of rotation of the coil is given by%
\begin{equation}
\varepsilon =\varepsilon _{\circ }+2\pi NAB\frac{1}{T}.
\end{equation}%
Here $\varepsilon _{\circ }$ is due to the background effect while $A$
stands for the area of the coil. We plot $\varepsilon $ versus $\frac{1}{T}$%
, which is a straight line with intercept $\varepsilon _{\circ },$ shown in
Fig. 3. The total magnetic field $B$ of the Earth is obtained as $B=0.41$
Gauss within the error limits.

We proceed next to determine the dip angle for the Earth's magnetic field.
For this purpose we mount our system such that it is rotated by $90^{\circ }$
to make the axis of rotation vertical (Fig. 2). We record the data for the
horizontal component $B_{h}$ and tabulate it in Tab. 2. We plot the induced
emf corresponding to $B_{h}$ in Fig. 4. This determines $B_{h}$ as $%
B_{h}=0.21$ gauss. The dip angle $\theta $ follows from 
\begin{equation}
\cos \theta =\frac{B_{h}}{B}
\end{equation}%
which turns out to be $\theta \tilde{=}59^{\circ }$ for Cyprus. The reason
that we obtained more than expected ($\approx 50^{\circ }$ for the
Mediterranean region) is due to the power lines in the surrounding.

\begin{acknowledgement}
We wish to thank the anonymous reviewer whose suggestions helped us to
improve the experiment.
\end{acknowledgement}

\bigskip \textbf{About the Authors}

Resat Akoglu, works as a Senior Physics Technician at EMU. E-mail:
resat.akoglu@emu.edu.tr. Mustafa Halilsoy is Professor and Chairman of
Physics Department at EMU. E-mail: mustafa.halilsoy@emu.edu.tr. S. Habib
Mazharimousavi is an assistant professor at EMU. E-mail:
habib.mazhari@emu.edu.tr

\textbf{Figure Captions}

Figure 1: Experiment setup when the axis of rotation is horizontal.

Figure 2: Experiment setup when the axis of rotation is vertical.

Figure 3: A plot of the amplitude of the induced emf versus the period of
rotation of the coil, when the axis of rotation is horizontal.

Figure 4: A plot of the amplitude of the induced emf versus the period of
rotation of the coil, when the axis of rotation is vertical.

\textbf{Tables}

Table 1:%
\begin{equation*}
\begin{tabular}{|l|l|l|}
\hline
& time for 10 turns $(s)$ & $emf\left( mV\right) $ \\ \hline
first & 03.90 & 16.2 \\ \hline
second & 04.95 & 14.2 \\ \hline
third & 06.00 & 12.5 \\ \hline
fourth & 06.60 & 11.3 \\ \hline
fifth & 07.50 & 09.5 \\ \hline
sixth & 08.20 & 08.5 \\ \hline
seventh & 12.00 & 05.0 \\ \hline
\end{tabular}%
\end{equation*}

Table 2:%
\begin{equation*}
\begin{tabular}{|l|l|l|}
\hline
& time for 10 turns $(s)$ & $emf\left( mV\right) $ \\ \hline
first & 04.00 & 08.5 \\ \hline
second & 05.20 & 07.7 \\ \hline
third & 05.90 & 06.5 \\ \hline
fourth & 06.50 & 06.0 \\ \hline
fifth & 07.00 & 05.5 \\ \hline
sixth & 08.50 & 04.7 \\ \hline
seventh & 12.50 & 03.2 \\ \hline
\end{tabular}%
\end{equation*}

\end{document}